\documentclass[prd,
twocolumn,
showpacs,preprintnumbers,amsmath,amssymb,
nofootinbib]{revtex4}

\usepackage{hyperref}
\usepackage{graphicx}
\usepackage{dcolumn}
\usepackage{bm} 
\usepackage{amsmath, amssymb, amsfonts}
\usepackage{multirow}
\newcommand\nn{\nonumber}

\newcommand{\Tr}{{\mathrm{\rm Tr}}}

\newcommand{\GEV}{\mbox{GeV}}

\newcommand{\bc}{\begin{center}} 
\newcommand{\ec}{\end{center}}   
\newcommand{\vecc}[1]{\mbox{\boldmath $#1$}}
\newcommand\ba{\begin{eqnarray}}
\newcommand\ea{\end{eqnarray}}
\newcommand{\be}{\begin{equation}}
\newcommand{\ee}{\end{equation}}

\begin{document}

\title{On the Measurement of Sachs Form Factors in Processes \\without and with Proton Spin Flip}

\author{M.~V.~Galynskii  \footnote{galynski@sosny.bas-net.by}}

\affiliation{Joint Institute for Power and Nuclear Research -- Sosny, BAS, 
 220109 Minsk, Belarus}
\begin{abstract}
The physical meaning of the decomposition of the Rosenbluth formula into two terms containing only squares of
Sachs form factors has been established. A new method has been proposed for their independent measurement
in the $e \vec{p} \to e\vec{p}$ elastic process when the initial proton at rest is fully polarized
along the direction of motion of the final proton.
\end{abstract}
\pacs{11.80.Cr, 13.40.Gp, 13.88.+e, 25.30.Bf}
\maketitle

{\bf \em {Introduction.}}---The electric ($G_{E}$) and magnetic ($G_{M}$) form factors
of the proton, the so-called Sachs form factors, in
elastic scattering of the electron on the proton have
been experimentally studied since the mid-1950s.
In the case of unpolarized electrons and protons, all experimental data on the behavior
of the form factors of the proton were obtained using the Rosenbluth formula
\cite{Rosen} for the differential cross section for the $ep\to ep$ elastic scattering
in the laboratory reference frame, where the initial proton is at rest. This formula
was obtained in the one-photon exchange and zero electron mass approximations and has the form
\ba
\label{Ros}
\frac{d\sigma} {d\Omega_e}= \frac{\alpha^2E_2\cos^2(\theta_e/2)}{4E_1^{\,3}\sin^4(\theta_e/2)}
\frac{1}{1+\tau_p} \left(G_E^{\,2} +\frac{\tau_p}{\varepsilon}G_M ^{\,2}\right).
\ea
Here, $\tau_p=Q^2/4M^2$,
$Q^2=-q^2=4E_1 E_2\sin^2(\theta_e/2)$
is the square of the momentum transferred to the proton and $M$ is the proton mass;
$E_1$, $E_2$, and $\theta_e$ are the energies of the initial and final electrons
and the angle of scattering of the electron in the laboratory reference
frame, respectively;
$\alpha=1/137$ is the fine structure constant;
and $\varepsilon=[1+2(1+\tau_p)\tan^2(\theta_e/2)]^{-1}$ is the degree
of the linear polarization of the virtual
photon \cite{Dombey,Rekalo74,GL97} varying in the region $0 \leqslant \varepsilon \leqslant 1$.

It was established with Eq. (\ref{Ros}) that the experimental dependence
of the Sachs form factors on $Q^2$ up to $Q^2\approx6$ \GEV$^2$ is dipole
and the ratio $R \equiv \mu_p G_E/G_M$ is approximately unity, $R \approx 1$,
where $\mu_p$=2.79 is the magnetic moment of the proton.

Akhiezer and Rekalo \cite{Rekalo74} proposed a method for measuring the ratio of the Sachs
form factors based on the transfer of polarization from the longitudinally
polarized initial electron to the final proton. This method involves the following
formula obtained in \cite{Rekalo74} for the ratio of form factors $G_E$ and $G_M$
in terms of the ratio of the degrees of the transverse, $P_t$, and longitudinal,
$P_l$, polarizations of the scattered proton:
\ba
\label{AxRek}
R \equiv  \frac{\mu_p\, G_E}{G_M}=-\frac{P_t}{P_l}\frac{E_1+E_2}{2M} \tan\left(\frac{\theta_e}{2}\right).
\ea
Precision experiments based on Eq.(\ref{AxRek}) were performed
at Jefferson Lab \cite{Jones00,Gay01,Gay02,Puckett12} in the range of
$0.5 \leqslant Q^2 (\GEV^2) \leqslant 8.5$. It was found that 
$R$ decreases linearly with increasing $Q^2$ in the range of
$0.5 \leqslant Q^2 (\GEV^2) \leqslant 5.6$ as
\begin{equation}
R =1-0.13\,(Q^2-0.04)\, \approx 1-Q^2/8\, ,
\label{linfit}
\end{equation}
which indicates that the form factor $G_E$ decreases faster than the form factor $G_M$.

Subsequent more accurate measurements of the ratio $R$ based on the Akhiezer--Rekalo
\cite{Puckett12} and Rosenbluth \cite{Qattan} methods confirmed the existence of discrepancies.
The current status of this problem is reviewed  \cite{ETG15}.

Since the results of the Sachs form factors measurements based on two experimental methods
are greatly differed it would be very important to measure them by another independent ways.

{\bf \em {
Cross section for the $e \vec p \to e \vec p$ process in the laboratory reference frame.}}---
In this work, we propose a new method for measuring the squares of the Sachs form factors
in the $e \vec p \to e \vec p$ elastic scattering, where $G_E^{\,2}$ and $G_M^{\,2}$ can be determined
independently from each other from direct measurements of cross sections for the process without
and with proton spin flip when the initial proton at rest is fully polarized along the direction
of motion of the scattered proton. This work is development of our work \cite{GKB2008}.

Let $s_{1}$ and $s_{2}$ be the spin 4-vectors of the initial and final protons
with 4-momenta $q_{1}$ and $q_{2}$, respectively, in an arbitrary reference frame.
The orthogonality ($s_{i} q_{i} = 0$) and normalization ($s_{i} ^{2} = - 1$) conditions
unambiguously provide the following expressions for the time ($s_{i0}$) and space ($\vecc s_i$)
components of these spin 4-vectors $s_i=(s_{i0}, \vecc s_i)$ in terms of their 4-velocities
$v_i=q_i/M$, $v_i=(v_{i0}, \vecc v_i)$:
\ba
s_i=(s_{i0}, \vecc s_i), \; s_{i0}=\vecc v_i\, \vecc c_i, \;
\vecc s_i =\vecc c_i + \frac{(\vecc c_i \vecc v_i)\,\vecc v_i}{1+v_{i0}}\;,
\label{spinv}
\ea
where $\vecc c_i$ ($\vecc c_i^{2}=1$) are the unit 3-vectors specifying the
directions of spin projections ($i=1, 2$).

In the laboratory reference frame, where $q_1=(M,\vecc 0)$ and $q_2=(q_{20}, \vecc q_2)$,
the directions of the spin projections $\vecc c_{1}$ and $\vecc c_{2}$ are chosen
such that they coincide with the direction of motion of the final proton:
\ba
\vecc c = \vecc c_{1} =\vecc c_{2}=\vecc n_2=  \vecc {q_2}/|\vecc q_2|\,.
\label{LSO}
\ea
Then, the spin 4-vectors $s_{1}$ and $s_{2}$ of the initial and
final protons, respectively, in the laboratory reference frame have the form
\ba
\label{DSB_LSO1}
s_1=(0,\vecc n_2 )\,, \; s_2= (|\vecc v_2|, v_{20}\, \vecc {n_2})\,, \,\vecc n_2=  \vecc {q_2}/|\vecc q_2|\,.
\ea

The proposed method is based on an expression for the differential cross section
for the $e \vec p \to e \vec p$ scattering in the laboratory reference frame
obtained in this work for the case where the initial and final protons are
polarized and have the common direction of the spin
projections $\vecc c$ (\ref{LSO}):
\ba
\label{RosPol}
\frac{d\sigma_{\delta_1, \delta_2}} {d\Omega_e}&=& \frac{\alpha^2E_2\cos^2(\theta_e/2)}
{4E_1^{\,3}\sin^4(\theta_e/2)} \frac{1}{1+\tau_p} \\ 
&\times& \left(\frac{1+\delta_1 \delta_2}{2}\, G_E^{\,2} 
+\frac{1-\delta_1 \delta_2}{2} \, \frac{\tau_p}{\varepsilon}\, G_M^{\,2}\right).\nn
\ea
Here $\delta_{1}$ and $\delta_{2}$ are the doubled spin projections of the initial and final
protons on the common direction of the spin projections $\vecc c$ (\ref{LSO}),  meanwhile,
$-1\leqslant \delta_{1,2}\leqslant 1$.

Two terms in Eq. (\ref{RosPol}) are contributions from transitions
without ($\sim G_E^{\,2}$) and with ($\sim G_M^{\,2}$) proton spin flip
in view of the polarization factors
\ba
\omega_{+}=(1 + \delta_1 \delta_2)/2, \, \,\omega_{-}=(1 -\delta_1 \delta_2)/2\,.
\label{omegi}
\ea

Indeed, according to Eq. (\ref{RosPol}), only the term containing $G_E^{\,2}$
contributes to the cross section for $e \vec p$ scattering
without proton spin flip ($\delta_1=1, \delta_2=1$) because $\omega_{+}=1$ and  $\omega_{-}=0$ in this case.

Only the term containing $G_M^{\,2}$ contributes to the cross section for $e \vec p$ scattering
with proton spin flip ($\delta_1=1$, $\delta_2=-1$) because $\omega_{+}=0$ and $\omega_{-}=1$
in this case. Consequently, the cross section the $e \vec{p} \to e \vec{p}$
process given by Eq. (\ref{RosPol}) can be represented as the sum of the cross
sections for processes without ($\sigma^{\uparrow\uparrow}$) and with
($\sigma^{\downarrow\uparrow}$)  proton spin flip:
\ba
\label{RosPol2}
\frac{d\sigma_{\delta_1, \delta_2}} {d\Omega_e}&=& 
\omega_{+} \sigma^{\uparrow\uparrow}+\omega_{-}\sigma^{\downarrow\uparrow},\\
\sigma^{\uparrow\uparrow}&=&\sigma_M \, G^2_E ,\;\;
\sigma^{\downarrow\uparrow}=\sigma_M \frac{\tau_p}{\varepsilon} \, G^2_M\,,
\label{RosPol2a}
\ea
where
\ba
\sigma_M= \frac{\alpha^2E_2\cos^2(\theta_e/2)}
{4E_1^{\,3}\sin^4(\theta_e/2)} \frac{1}{1+\tau_p}\,. \nn
\ea

The averaging and summation of Eq. (\ref{RosPol2}) over the polarizations
of the initial and final protons give the following representation for the Rosenbluth cross section
$\sigma_R=d\sigma / d\Omega_e$ specified by Eq. (\ref{Ros}):
\ba
\label{Ross}
\sigma_R
=\sigma^{\uparrow\uparrow} + \sigma^{\downarrow\uparrow}.
\ea

Consequently, the physical meaning of the decomposition of the Rosenbluth formula
into two terms containing only $G_E^{\,2}$ and $G_M^{\,2}$ is the sum of the cross sections
for processes without and with proton spin flip
when the initial proton at rest is fully polarized along the direction of motion of the final proton.
In this case, the electric $G_E$ and magnetic $G_M$ Sachs form factors determine
the contributions of the matrix elements of the proton current
for transitions of the proton without and with spin flip.

The validity of this treatment can be easily demonstrated as follows. The Rosenbluth
formula (\ref{Ros}) for unpolarized particles can be considered (see Eq. (\ref{MQED5}))
as the sum of cross sections for processes without and with spin flip of the initial proton,
which should be fully polarized along a certain direction determined by the kinematics
of the process. In the laboratory reference frame, the only separated direction is the direction
of motion of the scattered proton, and other separated directions are absent.

It is usually stated in the modern literature, in particular, in textbooks on the physics
of elementary particles (see \cite{XM}), that the Sachs form factors are simply
convenient because they allow the representation of the Rosenbluth formula in the simple and compact
form of the sum of two terms containing only $G^{\,2}_E$ and $G^{\,2}_M$. These formal reasons
for advantages of the Sachs form factors are included, in particular, in known
monographs \cite{AB,BLP}, are not criticized, and are reproduced
until now, e.g., in dissertation \cite{Paket2015}.

{\bf {\em Diagonal spin basis}.}---In the general case of the system of two particles with different momenta
$q_1=(q_{10}, \vecc {q_1})$ (before interaction) and $q_2=(q_{20}, \vecc {q_2})$
(after interaction), the possibility of the simultaneous projection of the spins on a single
common direction in an arbitrary reference frame is determined by
the three-dimensional vector \cite{FIF70}
\ba
\vecc a = \vecc q_{2}/q_{20} - \vecc q_{1}/q_{10} \,.
\label{os}
\ea

This result was obtained within the vector parameterization of the little Lorentz group $L_{q_1,q_2}$
common for two particles with 4-momenta $q_1$ and $q_2$ \cite{FIF70}
\ba
L_{q_1 q_2} q_1 =q_1, \; L_{q_1 q_2} q_2 =q_2\,. \nn
\ea
In particular, the initial and final protons in the $e \vec{p} \to e\vec{p}$
process can be considered as a system of two particles.

It is noteworthy that the group $L_{q_1,q_2}$ is implemented in a diagonal
spin basis \cite{Sik84}, where the spin 4-vectors of particles are expressed
in terms of their 4-momenta (4-velocities). The term "diagonal spin basis" is introduced
because the three-dimensional vector $\vecc a$ given by
Eq. (\ref{os}) is the difference of two vectors and is geometrically
the diagonal of a parallelogram.

According to Eq.~(\ref{os}), the common direction of the spin projections
in the rest system of the initial proton, where $\vecc {q_1}=0$, coincides with the direction of
motion of the final proton: $\vecc {a}=\vecc {n_2}=\vecc {q_2}/|\vecc {q_2}|$. This confirms
that the direction of motion of the final proton in the laboratory reference
frame is separate and can be the common direction of the spin projections.

It follows from Eq.~(\ref{os}) that the common direction of the spin projections
in the center of mass frame of colliding particles ($\vecc q_1+\vecc q_2=0$),
e.g., in the Breit system of the initial and final protons, where $q_1=(q_0, -\vecc q)$,
$q_2=(q_0, \vecc q)$, and  $q= q_2-q_1=(0, 2 \vecc q)$, as well as in the
laboratory reference frame, is the direction of motion of the final proton:
$\vecc {a}=\vecc {n_2}=\vecc {q_2}/|\vecc {q_2}| =\vecc {q}/|\vecc {q}|$, coinciding
with the direction of the momentum transfer.

In the diagonal spin basis, the spin 4-vectors $s_{1}$ and $s_{2}$
of the initial and final protons (fermions) with 4-velocities $v_{1}$ and $v_{2}$
($s_{i} v_{i} = 0$, $v_{i}^{2}=1 $, $s_{i}^{2}=-1$) have the form \cite{Sik84}

\ba
\label {DSB}
s_{1} = - \; \frac { (v_{1} v_{2}) v_{1} - v_{2}} {\sqrt{(v_{1}v_{2} )^{2} - 1 }} \; , \; \;
s_{2} =  \frac { ( v_{1} v_{2})v_{2} - v_{1}} {\sqrt{ ( v_{1}v_{2} )^{2} - 1 }} \, .
\ea

The spin 4-vectors given by Eqs. (\ref{DSB}) obviously do not change under transformations
of the little Lorentz group $L_{q_1,q_2}$ common for particles with 4-momenta $q_1$
and $q_2$. Therefore, they can be used to describe the spin states of the system of two
particles in an arbitrary reference frame by means of spin projections on a
single common direction of the 3-vector  given by  Eq. (\ref{os}). In particular,
the spin 4-vectors $s_{1}$ and $s_{2}$ specified by Eqs. (\ref{DSB})  have the form
of Eqs. (\ref{DSB_LSO1}) in the laboratory reference frame and correspond to the
directions of the spin projections given by Eqs. (\ref{LSO}).

The coincidence of the little Lorentz groups for particles with the 4-momenta $q_1$ and $q_2$
in the diagonal spin basis specified by Eqs. (\ref{DSB}) is responsible for a
number of remarkable properties of this basis. In particular, in the diagonal spin basis
specified by Eqs. (\ref{DSB}), the spin projection operators $\sigma_{1}$ and $\sigma_{2}$,
as well as the raising and lowering spin operators $\sigma_{1}^{\pm\delta}$ and
$\sigma_{2}^{\pm\delta}$, for the initial and final particles coincide with
each other \cite{GS98}:
\ba
&&\sigma = \sigma_{1} = \sigma_{2} =\gamma^{5} \hat{s_{1}} \hat{v_{1}} =
\gamma^{5} \hat{s_{2}} \hat{v_{2}}  = \gamma^{5} \hat{b}_{0} \hat{b}_{3} ,
 \label{spop6a} \\
&&\sigma^{\pm\delta} = \sigma_{1}^{\pm\delta} =\sigma_{2}^{\pm\delta}
= - 1/2 \,\gamma^{5}\, \hat{b}_{\pm\delta}\,,
 \label{spop6b}\\
&&\sigma u^{\delta}(q_{i}) = \delta u^{\delta}(q_{i}), \sigma^{\pm\delta} u^{\mp\delta}(q_{i})
= u^{\pm\delta}(q_{i}). \label{spop6c} 
\ea
Here, any 4-vector $\hat a$ is $\hat a = a_{\mu} \gamma^{\mu}$, $\gamma^5$
and $\gamma^{\mu}$ are the Dirac matrices, and $b_{\pm\delta} = b_{1} \pm i\, \delta b_{2}$
are the circular 4-vectors such that $b_{\pm\delta}^{~2}=0$, $\delta = \pm 1$.
In Eqs. (\ref{spop6a})  and (\ref{spop6b}), to construct the spin operators,
the following tetrad of orthonormalized 4-vectors $b_{A}$ $(A = 0, 1, 2, 3)$ is used:
\ba
\label{OBV}
&&b_0=q_+/\sqrt{q_+^2}\; , \; b_{3} = q_-/ \sqrt{-q_- ^2} \; \;, \\
&& (b_1)_{ \mu} = \varepsilon_{\mu \nu \kappa \sigma}b_0^{\nu}b_3^{\kappa}b_2^{\sigma},\;
(b_{2})_{\mu} = \varepsilon_{\mu \nu \kappa \sigma}q_1^{\nu}q_2^{\kappa} r^{\sigma}/\rho \, ,\nn
\ea
where $q_+=q_2+q_1$, $q_-=q_2-q_1$, $\varepsilon_{\mu\nu\kappa\sigma}$ is the Levi-Civita
tensor ($\varepsilon_{1230}=1$), $r$ is the 4-momentum of a particle involved
in the reaction different from $q_{1}$ and $q_{2}$ (e.g., 4-momentum of the initial
or final electron in the case of the $ e \vec p \to e \vec p$ process under consideration),
and $\rho$ is determined from the normalization conditions
$ b_{1}^{2} = b_{2}^{2} = b_{3}^{2}=-b_{0}^{2}=-1$. The coincidence of the
spin operators in the diagonal spin basis (\ref{DSB}) makes it possible to separate
in the covariant form the interactions without and with spin flip of particles involved in
the reaction and, thereby, to trace the dynamics of the spin interaction.


{\bf {\em Calculation of matrix elements of the QED processes
in the diagonal spin basis.}}---The amplitudes of QED processes in the scattering channel have the form
\be
M^{\pm\delta,\delta} = \overline {u}^{\pm \delta}(q_{2})
Q_{in} u^{\delta}(q_{1}) \, ,
\label{MQED}
\ee
where $u^{\delta}(q_{i})$=$u^{\delta}(q_{i},s_i)$ are the bispinors of the initial
and final states of fermions normalized as
$\overline {u}^{\delta}\, (q_{i}) u^{\delta}(q_{i}) = 2M$, $q_{i}^{2} = M^{2}$  $(i = 1, 2)$,
and $Q_{in}$ is the interaction operator.

The calculation of the matrix elements given by Eq. (\ref{MQED}) can be reduced
to the calculation of the trace of the product of operators:
\ba
\label{MQED1}
  M^{\pm\delta,\delta} = \Tr (P_{21}^{\pm\delta,\delta} Q_{in}),  \\
  P_{21}^ {\pm\delta,\delta} = u^{\delta}(q_{1})  \overline {u}^{\pm \delta}(q_{2}).
\label{P21pm}
\ea

The operators $P_{21}^{\pm \delta,\delta}$ (\ref{P21pm}) can be determined by several methods \cite{GS98,Cedrik18}.
In the used Bogush--Fedorov approach \cite{GS98}, in contrast to, e.g., \cite{Cedrik18}, the construction
of $P_{21}^{\pm \delta,\delta}$ is reduced to the determination of the
operators $T_{21}$ and $T_{12}$  
defined as
\ba
u^{\delta}(q_{2})=T_{21}u^{\delta}(q_{1}) , \,
u^{\delta}(q_{1})=T_{12}u^{\delta}(q_{2}),
\ea
such that $T_{12}=T_{21}^{-1}$, $T_{21}T_{12}=1$.

In the diagonal spin basis (\ref{DSB}), the operators $T_{21}$ and $T_{12}$ coincide
with each other \cite{GS98}:
\ba
T_{21} = T_{12} = \hat b_{0} \, .
\label{T120} \nn
\ea
As a result, the operator $P_{21}^{\delta,\delta}$ in Eqs. (\ref{P21pm}) is given by the
formula
\ba
P_{21}^{\delta,\delta} = u^{\delta}(q_{1}) \, \overline {u}^{\delta}(q_{1}) \,T_{12}
= \tau_{1}^{\delta}\, T_{12}=T_{12} \, \tau_{2}^{\delta}\,,
\label{P21pp}
\ea
where $\tau^{\delta}_i$=$u^{\delta}(q_{i}) \, \overline {u}^{\delta}(q_{i})$
are the projective operators for states of particles with 4-momenta $q_i$
and spin 4-vectors $s_i$ ($q_is_i=0, s_i^2=-1$, $i=1,2$) given by the formula
\ba
\tau^{\delta}_i = 1/2\, (\,\hat{q}_{i} + M) (\, 1 -  \delta \gamma^{5} \hat{s}_{i} ) \, .
\label{po1}
\ea

Relations (\ref{spop6a}) make it possible to represent the operators $\tau^{\delta}_{i}$ (\ref{po1})
in the diagonal spin basis in the form \cite{GS98}:
\ba
\tau^{\delta}_{i} = - 1/4 \; (\hat q_{i} + M ) \; \hat b_{\delta} \;  \hat b_{\delta}^{\ast} \; ,
\label{Proekoper}
\ea
where $b_{\delta}^{\ast} =b_{-\delta}  
= b_{1} - i \delta b_{2},  b_{\delta} b_{\delta}^{\ast}=-2,
b_{\delta}^2= b_{\delta}^{\ast 2}=0$.

The operator $P_{21}^{-\delta,\delta}$  in Eqs. (\ref{P21pm}) is reduced to the
product of the operators $\sigma^{+\delta}$ (\ref{spop6b})
and $P_{21}^{-\delta,-\delta}$ (\ref{P21pp}):
\ba
P_{21}^{-\delta,\delta} =\sigma^{+\delta}  P_{21}^{-\delta,-\delta}
= \sigma^{+\delta} \tau^{-\delta}_1 T_{12}\,  
 = \sigma^{+\delta} T_{12} \tau^{-\delta}_2.~~~
\label{P21pma}
\ea

As a result, the operators $P_{21}^{\pm\delta,\delta}$ in Eqs. (\ref{P21pm}) can be
represented in the compact form \cite{GS98}
\ba
\label{P21pp2}
P_{21}^{\delta,\delta} = ( \hat q_{1} + M) \, \hat b_{\delta} \,  \hat b_{0}
\; \hat b_{\delta}^{\ast} /4 \, ,  \\
P_{21}^{-\delta,\delta} = \delta (\hat q_{1} + M) \; \hat b_{\delta} \; \hat b_{3} /2\,.
\label{P21pm2}
\ea
These expressions allow the calculation of matrix elements for QED processes in the diagonal
spin basis that correspond to transitions without and with spin flip
for any interaction operator $Q_{in}$.

The matrix elements (\ref{MQED}) of QED processes can be
represented in the most general form
\be
M(\delta_1,\delta_2)
\equiv  M^{\pm\delta_2,\delta_1} = \overline {u}^{\pm
\delta_2}(q_{2}) Q_{in} u^{\delta_1}(q_{1}).
\label{MQED2}
\ee
For them hold true the following identities
\ba
M(\delta_1,\delta_2)= \omega_+  M^{+\delta_2,\delta_1} + \omega_-  M^{-\delta_2,\delta_1} .
\label{MQED3}
\ea

In view of the properties of the polarization factors
$\omega_{\pm}$ (\ref{omegi}) at $\delta_{1,2}=\pm1$
($\omega_{\pm}^2=\omega_{\pm}$, $\omega_{\pm}\omega_{\mp}=0$), we have
\ba
|M(\delta_1,\delta_2)|^2&=& \omega_+ | M^{+\delta_2,\delta_1}|^2 + \omega_-  |M^{-\delta_2,\delta_1}|^2 \,.
\label{MQED4}
\ea

In the $e \vec p \to e \vec p$ under consideration  process with unpolarized electrons,
all spin correlations in (\ref{MQED4}), excepting  those in $\omega_{\pm}$\,,
due to parity conservation in the electromagnetic interactions should be absent.
This means that $|M^{\pm\delta_2,\delta_1}|^2$ are independent of $\delta_1$ and $\delta_2$,
and the average value of the squared modules of matrix elements (\ref{MQED2})
summed over all polarizations has the form
\be
\overline{|M(\delta_1,\delta_2)|^2}= | M^{\uparrow\uparrow}|^2 +  |M^{\downarrow\uparrow}|^2 .
\label{MQED5}
\ee


{\bf \em {
Cross section for the $e \vec p \to e \vec p$ process in an arbitrary
reference frame.}}---In the one-photon exchange approximation, the matrix
element of the elastic process
\ba
e(p_1)+p(q_1,s_1) \to e(p_{2}) + p(q_2,s_2) 
\label{EPEP}
\ea
is the product of the electron and proton currents
\ba
\label{Mepep}
&& M_{ep\to ep} = 4\pi \alpha T / q^2\,, \\
&& T\equiv T^{\pm\delta,\delta }=(J_{e})^{\mu} ( J^{\pm \delta,\delta }_{p} )_{\mu} .
\label{T2pm}
\ea
The currents $(J_{e})^{\mu}$ and $(J^{\pm \delta,\delta }_{p})_{\mu}$ have the form
\ba
\label{Je}
 (J_{e})^{\mu} &=& \overline{u}(p_{2}) \gamma^{\mu} u(p_{1}) \,,  \\
\label{Jp}
 ( J^{\pm \delta,\delta }_{p} )_{\mu} &=&\overline{u}^{\pm \delta}(q_2)
\Gamma_{\mu}(q^{2}) u^{\delta}(q_1) \,, \\
\Gamma_{\mu}(q^{2}) &=& F_{1} \gamma_{\mu} + \frac{F_{2}} {4M}
( \; \hat q \gamma_{\mu} - \gamma_{\mu} \hat q \; ) \, ,
\label{Gamuepep}
\ea
where $u(p_{i})$ and $u(q_{i})$ are the bispinors of the electron
and proton with the 4-momenta $p_{i}$ and $q_{i}$, respectively;
$p_{i}^{2} = m^{2}$, $q_{i}^{2} = M^{2}$, $\overline{u}(p_{i})u(p_{i})=2m$,
$\overline{u}(q_{i}) u(q_{i})= 2M$ $(i = 1,2)$; $F_{1}$ and $F_{2}$
are the Dirac and Pauli form factors of the proton;
$q =q_{2}-q_{1}$ is the 4-momentum transferred to the proton; and $s_1$ and $s_2$
are the polarization 4-vectors of the initial and final protons, respectively.

The differential cross section for the $e \vec p \to e \vec p$ process has the form
\ba
\label{Modepep}
\frac {d \sigma}{d |t|}= \frac{ \pi \alpha^2 }{4 I^2 }\, \frac {|T|^2}{q^4} \; ,
\ea
where $ I^2=(p_1q_1)^2-m^2M^2, |t|=Q^2=-q^2$.

The matrix elements of the proton current (\ref{Jp}) calculated
by Eqs. (\ref{MQED1}), (\ref{P21pm}), (\ref{P21pp2}), and (\ref{P21pm2}) in the diagonal
spin basis (\ref{DSB}) have the form \cite{Sik84}
\ba
\label {Jepep-pp0}
&&( J^{\delta,\delta }_{p} )_{\mu} = 2 M \,G_{E} ( b_{0} )_{\mu} \, , \\
&&( J^{-\delta,\delta }_{p} )_{\mu}=- 2  M \,\delta \sqrt{\tau_p} \,G_{M} (b_{\delta })_{\mu}\, ,
\label {Jepep-pm0}
\ea
where
\ba
G_{E} = F_{1} -\tau_p \, F_{2} \, , \; G_{M} = F_{1} + F_{2}
\label {FFSep}
\ea
are the Sachs form factors, and $\tau_p=-q^2/4M^2$.

Consequently, the matrix elements of the proton current in the diagonal spin basis that correspond
to the proton transitions without and with spin flip given by Eqs. (\ref{Jepep-pp0})
and (\ref{Jepep-pm0}) are expressed only in terms of the Sachs form factors $G_{E}$ and $G_{M}$,
respectively. It is precisely because of this factorization of $G_E$ and $G_M$ that
Rosenbluth's formula is decomposed for the sum of two terms containing only $G_E^2$ and $G_M^2$,
which are responsible for the contributions of the transitions without and with spin flip
of the proton, respectively, when the directions of the spin projections of the protons
coincide with each other and have the form (\ref{os}).

With the use of the matrix elements (\ref{Jepep-pp0}) and (\ref{Jepep-pm0}),
the calculation of the quantities $|T|^2$ 
determining the cross section (\ref{Modepep}) for the $e \vec p \to e \vec p$
process is reduced to the calculation of traces:
\ba
\label{Tsqepep}
|T^{+ \delta,\delta}|^2& =&4M^2  G^{\,2}_E
\, \Tr ((\hat p_2 +m)\hat b_0(\hat p_1 +m)\hat b_0) /2\,, \\
|T^{ -\delta,\delta}|^2&=& 4M^2 \tau_p G^{\,2}_M
\Tr ((\hat p_2 +m)\hat b_{\delta}(\hat p_1 +m) \hat b^{\ast}_{\delta})/ 2.~~~~
\label{Tsqepep2}
\ea
The simple calculations give
\ba
\label{Wep pp}
|T^{+ \delta,\delta}|^2&=& \frac{G^{\,2}_E}{1+\tau_p}\,  Y_1 ,\;\;
|T^{- \delta,\delta}|^2 =\frac{\, \tau_p \, G^{\,2}_M} {1+\tau_p} \, Y_2\,,\\
Y_1&=&(p_+q_+)^2+q_+^2q_-^2\,,\\
Y_{2}&=&(p_+ q_+)^2-q_+^2(q_-^2+4 m^2)\,,
\label{Wep pm}
\ea
where $p_+=p_1+p_2$, $q_+=q_1+q_2$, and  $q_-=q_2-q_1=q$.

First, the quantities $|T^{\pm \delta,\delta}|^2$  given by Eqs. (\ref{Wep pp}) are
independent of the polarization of protons in agreement with the above discussion.
Second, denominators in
Eqs. (\ref{Wep pp}) are due to the normalization of the 4-vector
$b_0$  in Eqs. (\ref{OBV}) and to the relation $q_+^2=4M^2(1+\tau_p)$.

The quantity $|T|^2$ determining the cross section
(\ref{Modepep}) for the $e \vec p \to e \vec  p $  process in the diagonal spin basis
(\ref{DSB}) has the form (see Eq. (\ref{MQED4}))
\ba
|T^{\delta_1, \delta_2}|^2 =\omega_+ |T^{+ \delta,\delta}|^2
+\omega_- |T^{- \delta,\delta}|^2\,.
\label{Td}
\ea
The calculation of the quantity $|T^{\delta_1, \delta_2}|^2$ in the diagonal
spin basis by standard methods \cite{AB,BLP} also gives a
result coinciding with Eq. (\ref{Td}).

The resulting differential cross section for the $e \vec{p} \to e \vec{p}$
process in the diagonal spin basis in an arbitrary reference frame has the form
\ba
\label{crepep1}
\frac {d \sigma_{\delta_1, \delta_2}}{d |t|}= \frac{ \pi \alpha^2 }{4 I^2 (1+\tau_p)}
\left(  \omega_+ G_{E}^{2}  Y_{1} + \omega_-
\tau_p \, G_{M}^{2}  Y_{2}  \right) \frac{1}{q^4 }.~
\ea

The multiplication of Eq. (\ref{crepep1}) with $\delta_2=0$ by a factor
of 2 gives the following cross section for the $ep \to ep$
process, where all particles are unpolarized,
in an arbitrary reference frame:
\ba
\label{crepep}
\frac {d \sigma}{d |t|} = \frac{ \pi \alpha^2 }{4 I^2\, (1+\tau_p)}  \, ( \, G_{E}^{2} \; Y_{1}
+ \tau_p \; G_{M}^{2} \; Y_{2} \, )\, \frac{1}{q^{4}} \, .
\ea
This expression coincides with Eq. (34.3.3) from \cite{AB}.

It is convenient to represent the above expressions for $Y_{1}$ and $Y_{2}$
in terms of the Mandelstam variables:
\ba
s=(p_1+q_1)^2, \, t=(q_2-q_1)^2, \, u=(q_2-p_1)^2. \nn
\ea
Inverting the relation, we have
\ba
p_+q_+=s-u, \, q_+^2=4M^2-t, \, q_-^2=t, \, \tau_p=-t/4M^2, \nn
\ea
whence
\ba
&& Y_1=(s-u)^2+(4M^2-t)\,t, \; \;\\
&& Y_2=(s-u)^2-(4M^2-t)(t+4m^2)\,.
\ea
The quantity $4I^2$ in Eq. (\ref{Modepep}) has the form
\ba
4I^2=(s-(M+m)^2)(s-(M-m)^2)=\lambda(s,m^2,M^2), \nn
\ea
where $\lambda(s,m^2,M^2)$ is the K\"{a}ll\'{e}n function.

It is noteworthy that, if the mass of the electron is
neglected, Eq. (\ref{crepep1}) in the laboratory reference frame
has the form of Eq. (\ref{RosPol}), and Eq. (\ref{crepep}) written in variables
$s, t$, and $u$, coincides with Eq. (139.4) from \cite{BLP}.

\newpage
{\bf {\em The helicity amplitudes of the $e \vec p \to e \vec p$ process
in the Breit system.}}--- Information reported in the literature is sufficient
to understand the physical meaning not only
the decomposition of the Rosenbluth formula (\ref{Ros}) but also of the Sachs form factors.
Such an understanding could be based on exercise (8.7) in \cite{XM},
where the explicit form of the helicity
amplitudes of the proton current $J^{\pm \lambda, \lambda}_{p}$ is presented in
the Breit system, where $q_1=(q_0, -\vecc q)$ and $q_2=(q_0, \vecc q)$,
i.e., $q=q_2-q_1= (0, 2 \vecc q)$.
The amplitudes presented in \cite{XM} have the form
\ba
\label{tokHM}
&&(J^{-\lambda,\lambda}_p)_{\mu}=2 M\, G_{E} (b_{0})_{\mu},\\
&&(J^{\lambda,\lambda}_p)_{\mu}=  -2 \lambda \,|\vecc q|\, G_{M} (b_{\lambda})_{\mu},
\label{tokHM2}
\ea
where $M$ is the proton mass, $2|\vecc q|=\sqrt{Q^2}$, $Q^2=-q^2$,
$b_{\lambda}$ is the circular 4-vector, $b_{\lambda} = b_{1} + i \lambda b_{2}$,  $ \lambda=\pm 1$,
\ba
\label{BAA}
&& (b_{0})^{\mu}=(1,0,0,0), (b_1)^{\mu}=(0,1,0,0), \\
&&(b_2)^{\mu}=(0,0,1,0), (b_3)^{\mu}=(0,0,0,1)\,.\nn
\ea

\begin{figure}[h!]
\centerline{
\includegraphics[scale=0.80]{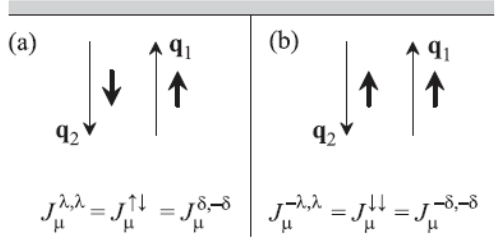}}
\vspace{-0.5cm}
\caption{
Breit system for the initial and final protons, where
$\vecc q_1 +\vecc q_2=0$. Long (short) arrows denote the momenta
(spins) of protons. Panel (a) corresponds to a transition
with helicity conservation ($\lambda_2=\lambda_1=+1$) at which the
proton spin is flipped. Panel (b) corresponds to a transition
of a proton with the change in the sign of helicity
($\lambda_1=+1, \lambda_2=-1$) at which the proton spin is not flipped.
}
\label{BS}
\end{figure}

Since the directions of momenta of the initial and final protons in the Breit system
are opposite to each other, a transition of the proton with helicity sign conservation
is a transition with proton spin flip
($J_{\mu}^{\lambda, \lambda}=J_{\mu}^{\uparrow \downarrow } =J_{\mu}^{ \delta, -\delta}$),
and a transition of the proton with a change in the helicity sign is a transition without proton
spin flip ($J_{\mu}^{-\lambda, \lambda}= J_{\mu}^{\downarrow \downarrow}=J_{\mu}^{-\delta,\, -\delta}$).
This is clearly seen in Fig. \ref{BS}.

Transitions with helicity conservation are often called transitions without spin flip.
The example of the Breit system shows that this statement can be erroneous.

According to the above presentation, the matrix elements given by Eqs. (\ref{tokHM}) and (\ref{tokHM2})
can be represented in terms of the amplitudes $J_{p}^{\pm \delta, \delta}$ ($\delta=\pm 1$)
corresponding to transitions of the proton without and with spin flip for the case where
the spins of the initial and final protons are projected on a single common direction coinciding
with the direction of motion of the final proton in the Breit system:
\ba
\label {BSO_DSB}
&&(J_p^{\delta,\delta})_{\mu} = 2 M \, G_{E} (b_{0})_{\mu},\\
&&(J_p^{-\delta,\delta })_{\mu}=-2 \delta  M \, \sqrt{\tau_p}\, G_{M} (b_{\delta })_{\mu}.
\label {BSO_DSB2}
\ea

First, the matrix elements specified by Eq.~(\ref{BSO_DSB}) and (\ref{BSO_DSB2}) are
more convenient than the helicity amplitudes given by
Eqs.~(\ref{tokHM}) and (\ref{tokHM2}), e.g., in the case of passage from the Breit
system to the laboratory reference frame, where the
notion of helicity is inapplicable for the initial proton.
Second, the explicit form of the matrix elements specified
by Eq.~(\ref{BSO_DSB}) and  (\ref{BSO_DSB2}) coincides with the matrix elements
given by Eqs.~(\ref{Jepep-pp0}) and (\ref{Jepep-pm0}) in the diagonal spin basis
(which are valid in an arbitrary reference frame);
amplitudes (\ref{BSO_DSB}) and  (\ref{BSO_DSB2}) are particular cases of matrix elements
(\ref{Jepep-pp0}) and (\ref{Jepep-pm0}). To demonstrate this, it is sufficient to
verify that the tetrad of 4-vectors (\ref{OBV}) is transformed in
the Breit system to the tetrad of unit 4-vectors (\ref{BAA}).
We consider only the 4-vectors $b_0$ and $b_3$ in Eqs.~(\ref{OBV}) and
(\ref{BAA}). Indeed, since the 4-vectors $b_0$ and $b_3$ in (\ref{OBV}) are
the sum $q_+/\sqrt{q_+^2}$ and difference $q_-/\sqrt {-q_-^2}$ of 4-momenta
of protons normalized to unity, in the Breit system with the third axis along
the direction of motion of the final proton, this sum and difference have the form
$q_+=q_1+q_2=(2q_0, 0,0,0)$ and $q_-=q_2-q_1=(0,0,0,2|\vecc q|)$.
Being normalized, the 4-vectors $b_0$ and $b_3$
in Eqs.~(\ref{OBV}) are transformed to the unit 4-vectors $b_0=(1,0,0,0)$
and $b_3=(0,0,0,1)$ in the Breit system. It can be demonstrated similarly that
the 4-vectors $b_1$ and $b_2$ in Eqs.~(\ref{OBV}) are transformed to the corresponding
4-vectors of tetrad (\ref{BAA}) in the Breit system.

For the inverse transformation of helicity (\ref{tokHM}) and  (\ref{tokHM2}) or
diagonal (\ref{BSO_DSB}) and (\ref{BSO_DSB2})   amplitudes in the Breit system to matrix
elements (\ref{Jepep-pp0}) and (\ref{Jepep-pm0}) of the proton current in the
diagonal spin basis, the tetrad of vectors (\ref{BAA}) appearing
in the amplitudes of the proton current $J_{\mu}^{\pm\delta, \delta}$ (\ref{BSO_DSB})  and (\ref{BSO_DSB2})
should be expressed in terms of the 4-momenta of particles involved in the reaction using the algorithm of
construction of the tetrad of 4-vectors (\ref{OBV}). In this case, it is not necessary
to apply Lorentz transformations to obtain expressions valid in an arbitrary reference frame.

Thus, the example of consideration of matrix elements of the proton current in the Breit
system in \cite{XM} can help to understand the physical meaning of
both the decomposition of the Rosenbluth formula and the Sachs form factors.
Unfortunately, readers of \cite{XM} did not notice this assistance.

Since only the time and space components of the 4-vector $J^{-\lambda,\lambda}_{\mu}$
and $J^{\lambda,\lambda}_{\mu}$ in Eqs.~(\ref{tokHM}) and (\ref{tokHM2}) are nonzero,
respectively, these 4-vectors are erroneously joined in \cite{XM} into a single 4-vector,
i.e., written in the form
\ba
(J_p)_{\mu}=(j_0, \vecc j), \, j_0=(J^{-\lambda,\lambda})_0, \, \vecc j=\vecc {J}^{\lambda,\lambda}.\nn
\ea

The identities in Eqs. (\ref{MQED3}) allow correcting this error:
\ba
(J_p)_{\mu} 
= \omega_+  (J_p^{+\delta,\delta})_{\mu} + \omega_-  (J_p^{-\delta,\delta})_{\mu} \,,
\label{Jpdd}
\ea
where $(J_{p}^{\pm\delta,\delta})_{\mu}$ are the matrix elements of the proton
current (\ref{Jepep-pp0}) and (\ref{Jepep-pm0}) in the diagonal spin basis.

As a result, for the symmetric parts of the hadron
($H_{\mu \nu}=(J_p)_{\mu}(J_p^{\ast})_{\nu}$) and lepton ($L^{\mu \nu}=(J_e)^{\mu}(J_e^{\ast})^{\nu}$)
tensors we have
\ba
\label{Hmunu}
&&H_{\mu \nu}
=\omega_+ H_{\mu \nu}^{\delta, \delta} + \omega_- H_{\mu \nu}^{-\delta, \delta},\\
\label{Hpp}
&&H_{\mu \nu}^{\delta, \delta}=4M^2\,G_E^2\,\frac{(q_+)_{\mu} (q_+)_{\nu}}{q_+^2}\,, \\
\label{Hpm}
&&H_{\mu \nu}^{-\delta, \delta}=4M^2 \tau_p \,G_M^2\,
\left (- g_{\mu \nu}  + \frac{  (q_+)_{\mu} (q_+)_{\nu}}{q_+^2}  \right) , \\
\label{Lmunu}
&&L^{\mu \nu} 
=p_+^{\mu} p_+^{\nu} + q^2 g^{\mu \nu},
\ea
where $g_{\mu \nu} =\rm {diag} (1, -1, -1, -1)$ is the metric tensor in Minkowski space.

Thus, the hadron tensor $H_{\mu \nu}$ (\ref{Hmunu}) in the diagonal
spin basis is naturally separated into contributions corresponding
to transitions without, Eq.~(\ref{Hpp}), and with,
Eq.~(\ref{Hpm}), proton spin flip, which are simultaneously the
longitudinal ($H_{\mu \nu}^{\delta, \delta}=H_{\mu \nu}^L$) and transverse
($H_{\mu \nu}^{-\delta, \delta}=H_{\mu \nu}^T$)
contributions, respectively.

The product of the tensors $H_{\mu \nu}$ and $L^{\mu \nu}$ gives
Eq. (\ref{Td}) for $|T^{\delta_1, \delta_2}|^2$.

Tensor $H_{\mu \nu}$ for unpolarized protons has the form
\ba
\overline{H}_{\mu \nu} = H_{\mu \nu}^{L} + H_{\mu \nu}^{T}\,.
\label{Hppn}
\ea

{\bf {\em Conclusions.}}---The differential cross section for the $e \vec p \to e \vec p $ process
in the diagonal spin basis has been calculated in
the Born approximation in an arbitrary reference
frame. Expression (\ref{RosPol}) obtained for the cross section in
the laboratory reference frame can be used to measure
the squares of the Sachs form factors, $G_E^{\,2}$  and $G_M^{\,2}$, in
processes without and with spin flip for the case where
the initial proton is fully polarized along the direction
of motion of the final proton. In the asymptotic limit
of large $Q^2$ values, where $\tau_p \gg 1$, the contribution to
cross section (\ref{crepep1}) comes only from transitions with
proton spin flip where helicity is conserved. To understand
this, it is sufficient to pass from an arbitrary reference
frame to the Breit system and to use Fig. \ref{BS}. It is
noteworthy that the conclusion of spin flip in processes
with helicity conservation is valid not only for
protons but also for point electrons at $\tau_e \gg 1$.

\end{document}